\documentclass[11pt,twoside]{article}
\usepackage{asp2010}
\usepackage{graphicx}
\usepackage{grffile}
\usepackage{natbib}

\resetcounters

\begin{document}
\label{authorguide}

\allowtitlefootnote

\title{Dynamical Friction in Cuspy Galaxies}
\author{M. Arca-Sedda, R. Capuzzo-Dolcetta
\affil{Dep. of Physics, ``Sapienza'', Universit\`a di Roma, Piazzale Aldo Moro 2, 00185, Rome, Italy}}






\begin{abstract}
It is well known that a large fraction of galaxies have cuspy luminosity profiles in their central regions, at least within the observational resolution. In such cases, the often used, simplified, local approximation for the dynamical friction braking classical term fails when the massive satellite moves through the inner parts of the galaxy, although the scattering integral still converges for phase space  distribution singularities that are not too sharp. 
Here we present preliminary results of our work aiming at finding better and more reliable results from the integration of motion of massive objects (globular clusters) in galaxies where the density diverges to the center in a power law form, with exponent greater than $-2$.
\end{abstract}

\section{Dynamical friction}
The frictional deceleration of the motion of a $test$ particle of mass $M$ placed, at time $t$, at ${\mathbf r}$ is given by the cumulative effect of a background of $target$ stars of mass $m$  at varying the impact vector ${\mathbf b}$, which can be expressed as 

\begin{equation}
\frac{d{\mathbf v}_M}{dt} = -{{2m}\over {M+m}}
\int_{{\mathbf b}}\int_{{\mathbf v}_m}
f(\mathbf{r+b,v}_m)
\frac{{\mathbf v}_M-{\mathbf v}_m}{1+\frac{b^2
|{\mathbf v}_m-{\mathbf v}_M|^4}{G^2(M+m)^2}}{|{\mathbf v}_M-{\mathbf v}_m| \over b}
d^3\mathbf{v_m}d^3\mathbf{b},
\label{dfcorrect1}
\end{equation}
where $f(\mathbf{r+b,v}_m)$ is the stellar background distribution function (DF) at position $\mathbf{r}'=\mathbf{r+b}$.
Even for a DF inspace and velocity spherical simmetry, i.e. $f(r',v_m)$, the integral in Eq.(\ref{dfcorrect1}) is very complicated whenever the test particle is out of the origin ($r\neq 0$).

\begin{figure}
{\includegraphics[width=6.5cm]{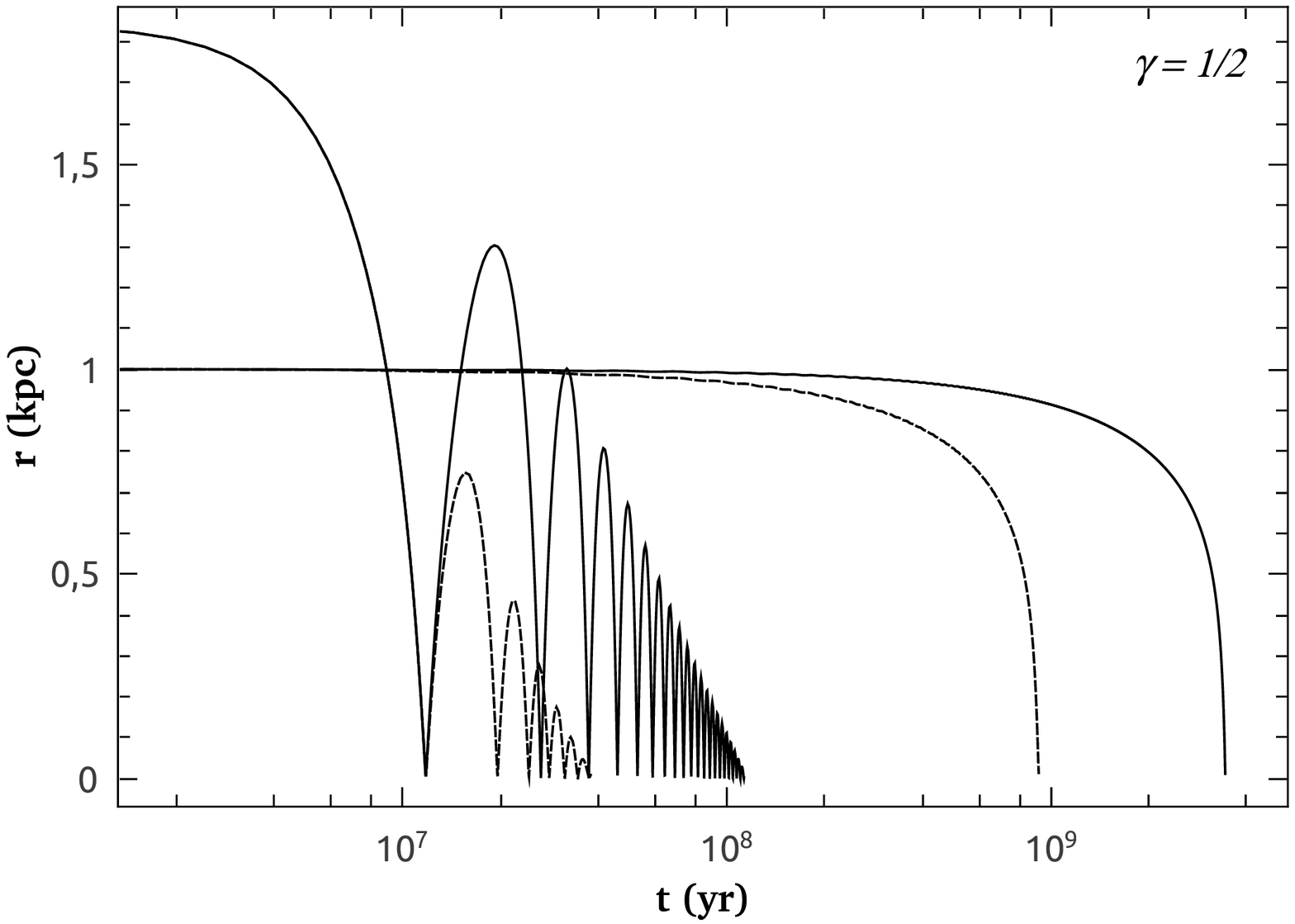}}
\hspace{0.5cm}
{\includegraphics[width=6.5cm]{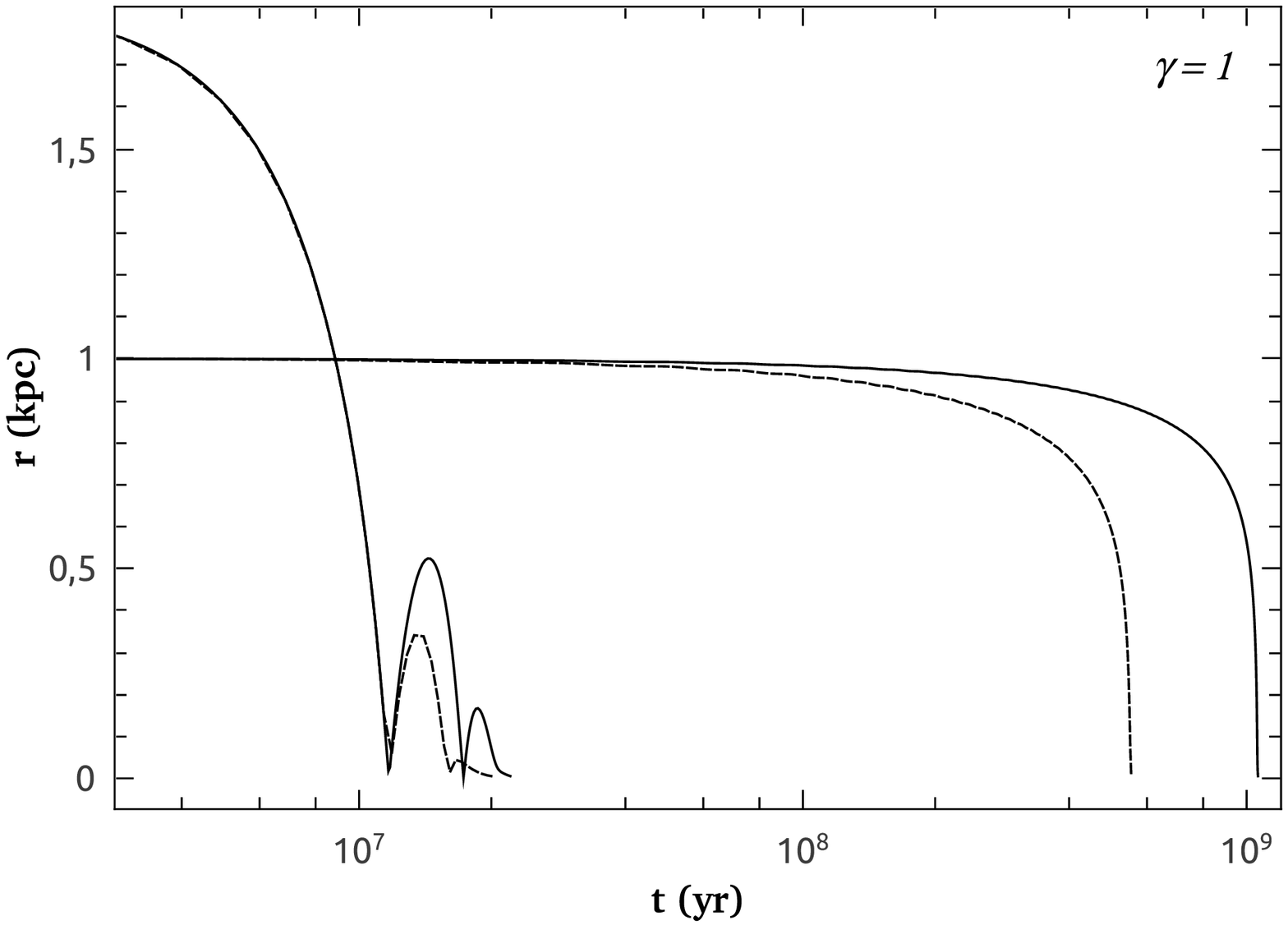}}
\caption{The time evolution of the distance to the galactic center of a massive test object 
in the case of initially circular orbits (regular curves) and radial orbits (oscillating curves) of same initial orbital energy, 
for two values of $\gamma$, as labeled, and two values of the object mass 
(upper curves refer to $M=10^6$ M$_\odot$. lower curves to $M=4 \times10^6$ M$_\odot$).}
\label{fig1}
\end{figure}

To overcome, partially, this problem, the common approach consists in
taking out of the integral in Eq. (\ref{dfcorrect1}) the DF, assumed as independent of ${\mathbf b}$ and isotropical in the velocity space, thus yielding to what is often referred as {\it local} approximation formula \cite{bt}

\begin{equation}
\left({ \frac{d{\mathbf v}_M}{dt} }\right)_{loc}= -4 \pi G^2 M \rho(r) F(r,v_M) \ln\Lambda 
{{{\bf v}_M} \over {v_M^3}},
\label{ads}
\end{equation}
the function $0\leq F(r,v_M)\leq 1$ being the local fraction of target stars slower than the test object ($v_m \leq v_M$). 
In the local approximation, the mass density, $\rho$, of field stars in Eq. (\ref{ads}) is assumed at its
{\it local} value, i.e. that evaluated at the position (${\mathbf r}$) of the test particle; $\ln\Lambda$ is the usual Coulomb's logarithm.

Now, whenever the test particle is significantly off center with respect to the stellar system (star cluster, galaxy, etc.) wherein it moves,  the local expression gives (for symmetry reasons) a quite acceptable approximation; on the contrary, it loses its validity in the neighbourhood of the centre. In this case the local approximation is clearly an overestimate of the actual dynamical friction, 
because it corresponds to weighting the contribution of the 
gravitational encounters with objects at a given distance from the test particle not with the, correct, density of target stars at that distance but, rather, with the density of targets
evaluated at the location of the test particle itself, that is maximum at the origin of any self-gravitating system.
This overestimate is particularly relevant when dealing with {\it cuspy} galaxies, where the spatial density of stars diverges at the galactic center, making the local approximation unviable. \\
To solve this problem, we follow the approach presented in \citet{vicetal} and deeply discussed in \citet{cdas11}), which is based on taking into account that at the origin ($r=0$) and in the hypothesis of isotropic DF 
the integral in Eq. (\ref{dfcorrect1}) simplifies into the form 

\begin{equation}
\left({ \frac{d{\mathbf v}_M}{dt} }\right)_{cen} = -{{8\pi m}\over {M+m}}
\int_{b_{min}}^{b_{max}}\int_{{\mathbf v}_m}
f(b,\mathbf{v}_m)
\frac{{\mathbf v}_M-{\mathbf v}_m}{1+\frac{b^2
|{\mathbf v}_m-{\mathbf v}_M|^4}{G^2(M+m)^2}}|{\mathbf v}_M-{\mathbf v}_m|b d^3\mathbf{v_m} db.
\label{dfcorrect2}
\end{equation}
which can be performed numerically, although with the care due to the double singularity in phase space. 
Once that the integral in Eq. (\ref{dfcorrect2}) is evaluated, an estimate of the dynamical friction deceleration along the test object trajectory is given by an interpolation between the local approximation as given by Eq. \ref{ads} and the central value given by Eq. (\ref{dfcorrect2}), in the form

\begin{equation}
\frac{d{\mathbf v}_M}{dt} = p(r)\left({ \frac{d{\mathbf v}_M}{dt} }\right)_{cen}+
[1-p(r)] \left({ \frac{d{\mathbf v}_M}{dt}
}\right)_{loc} ,
\label{dfinterp}
\end{equation}
where the interpolation function $0\leq p(r)\leq 1$ is such that $p(0)=1$. We found $p(r)=e^{-r/R}$ (where $R$ is a proper length scale) as a good interpolation formula.
We have applied this to the orbital evolution of massive objects in a spherical potential resembling that of the Milky Way (disk excluded), as given by the sum of three components: i) a cuspy inner bulge, represented as  a 
Dehnen's $\gamma$-model \citep{deh}, ii) an extended bulge-halo represented as a Plummer sphere \citep{plu} and, iii) a dark matter halo in the Navarro, Frenk and White form \citep{nfw}.

\section{Some results}
The comparison between the efficiency of dynamical friction on circular and radial orbits of same energy is particularly important 
because they constitute (for any given orbital energy, in a spherical potential) the lower and upper limits of the expected dynamical braking.
In Fig. \ref{fig1} we show, for two values of $\gamma$ in the Dehnen's model, the time evolution of the distance to the center for circular orbits starting from $r=1$ kpc from the galactic center 
and for the radial orbits of same energy, for 2 values of the test object mass, $M=10^6$ M$_\odot$ and $M=4\times 10^6$ M$_\odot$. The ratio of the radial to circular dynamical friction decay times, i.e. the time needed to lose almost all the initial orbital energy, is about $0.04$ for both  $\gamma = 1/2$ and $\gamma = 1$, indicating how efficient is dynamical frictional braking through galactic, cuspy, centers and, so, how important is its careful treatment.

\bibliography{ArcaSedda_M}

\begin{thebibliography}{}
\expandafter\ifx\csname natexlab\endcsname\relax\def\natexlab#1{#1}\fi
\expandafter\ifx\csname url\endcsname\relax
  \def\url#1{\texttt{#1}}\fi
\expandafter\ifx\csname urlprefix\endcsname\relax\def\urlprefix{URL }\fi
\providecommand{\eprint}[2][]{\url{#2}}

\bibitem[{Binney \& Tremaine(1987)}]{bt}
Binney, J., \& Tremaine, S. 1987, Galactic Dynamics (Princeton, New Jersey:
  Princeton Univ. Press), 2nd ed.

\bibitem[{Capuzzo-Dolcetta \& Arca-Sedda(2011)}]{cdas11}
Capuzzo-Dolcetta, R., \& Arca-Sedda, M. 2011, in preparation

\bibitem[{{Dehnen}(1993)}]{deh}
{Dehnen}, W. 1993, \mnras, 265, 250

\bibitem[{{Navarro} et~al.(1996){Navarro}, {Frenk}, \& {White}}]{nfw}
{Navarro}, J., {Frenk}, C., \& {White}, S. 1996, Astrophys.J., 462, 563

\bibitem[{Plummer(1911)}]{plu}
Plummer, H. 1911, MNRAS, 71, 460

\bibitem[{Vicari et~al.(2007)Vicari, Capuzzo-Dolcetta, \& Merritt}]{vicetal}
Vicari, A., Capuzzo-Dolcetta, R., \& Merritt, D. 2007, Astrophys. J., 662, 797

\end{thebibliography}
\bibliographystyle{asp2010}

\end{document}